\documentclass[aps,pra]{revtex4}
\usepackage{graphicx}

\begin{document}

\title{Improving quantum dense key distribution \\
\thanks{*Email: zhangzj@wipm.ac.cn }}

\author{Z. J. Zhang and Z. X. Man \\
{\normalsize Wuhan Institute of Physics and Mathematics, Chinese
Academy of Sciences, Wuhan 430071, China } \\
{\normalsize *Email: zhangzj@wipm.ac.cn }}

\date{\today}
\maketitle
\begin{minipage}{380pt}
The capacity of the quantum dense key distribution (QDKD) [Phys.
Rev. A69, 032310 (2004)] is doubled by introducing the dense
coding. The security of the improved QDKD against eavesdropping is
pointed out to be easily proven.  In both the original QDKD and
the present improved QDKD, a strategy to double the efficiency of
generating the secret key with given length is proposed. In
addition, we point out a leak of security of the original QDKD and fix it. \\

PACS Number(s): 03.67.Dd, 03.65.Hk\\
\end{minipage}

Recently, Bostroem and Felbinger [1] have proposed a deterministic
secure direct communication protocol, where the original idea that
the secure information is encoded by a local operation on a photon
of the Einstein-Podolsky-Rosen (EPR) photon pair is presented.
However, in favor of a secure transmission, in their protocol they
abandon {\it the dense coding} feature. It is known that {\it the
quantum dense coding} (QDC, a different concept from {\it the
dense coding}) by using local operations on one particle of EPR
pair shared by two parties doubles basically the capacity of
transmission of a classical channel [2]. So very recently, I. P.
Degiovanni et al [3] have proposed a deterministic quantum dense
key distribution (QDKD) by including the advantages of the QDC by
using local unitary operations and the quantum key distribution
(QKD) in [1] in generating shared secret keys and enhancing
transmission capacity. In fact, the transmission capacity of the
QKD in Ref. [1] can be doubled by introducing {\it the dense
coding} and the security has been proven [4]. Therefore, the QDKD
proposed very recently by I. P. Degiovanni et al [3] can be
further improved. We briefly report it in this paper. In addition,
we point out a leak of security in the original QDKD and fix it.

Our present improved QDKD is depicted simply as follows. The four
Bell states of photons $h$ and $t$ are $|\Psi^{\pm}_{ht}
\rangle=(1/\sqrt{2})(|0\rangle_h|1\rangle_t\pm|1\rangle_h|0\rangle_t)$
and
$|\Phi^{\pm}_{ht}\rangle=(1/\sqrt{2})(|0\rangle_h|0\rangle_t\pm|1\rangle_h|1\rangle_t)$.
Alice prepares an entangled photon pair in the state $|\Psi^+_{ht}
\rangle=(1/\sqrt{2})(|0\rangle_h|1\rangle_t+|1\rangle_h|0\rangle_t)$,
and she stores photon $h$ in her lab, whereas she performs
randomly with one of the local unitary operations
$u_0=|0\rangle\langle0|+|1\rangle\langle1|$,
$u_1=|0\rangle\langle0|-|1\rangle\langle1|$,
$u_2=|0\rangle\langle1|+|1\rangle\langle0|$,
$u_3=|0\rangle\langle1|-|1\rangle\langle0|$ on photon $t$ and then
sends it to Bob. After this the initial state $|\Psi^+_{ht}
\rangle$ changes into $|\Psi^+_{ht} \rangle=u_0|\Psi^+_{ht}
\rangle$, $|\Psi^-_{ht} \rangle=u_1|\Psi^+_{ht} \rangle$, $|\Phi^+
_{ht}\rangle=u_2|\Psi^+_{ht} \rangle$, or $|\Phi^-_{ht}
\rangle=u_3|\Psi^+_{ht} \rangle$, so Alice's random selection of
the unitary operations can be taken as a two-bit encoding of her
secret key on the EPR pair, i.e., $u_0$ as '00', $u_1$ as '01',
$u_2$ as '10' and $u_3$ as '11'. When Bob receives the photon $t$,
he can have two choices. One choice is that he performs a
measurement on the photon $t$ by choosing randomly either the base
$\{|0\rangle_t,|1 \rangle_t \}$ or the base $\{(|0\rangle_t+|1
\rangle_t)/\sqrt{2}, (|0\rangle_t-|1 \rangle_t)/\sqrt{2} \}$ and
then he publicly announces the base he used and his measurement
result. After Bob's public announcement Alice performs her
measurement on the photon $h$ by choosing the same base as Bob.
Since she knows which unitary operation she has performed on the
photon $t$, in the case of $u_0$ or $u_1$ ( $u_2$ or $u_3$), if
she finds that her measurement result is correlated
(anticorrelated) with Bob's measurement result, she tells Bob that
Eve is in the line and then their transmission is aborted.
Otherwise, their transmission continues. The other choice of Bob
is that he encodes his bits via performing one of the four local
unitary operation on the photon $t$ and then sends it back to
Alice. When Alice receives the back photon $t$, she performs the
Bell state measurement on the photons $h$ and $t$ and announce
publicly her measurement result. Then both Bob (Alice) can know
the exact local unitary operation performed by Alice (Bob) on the
photon $t$ (see table 1).  Thus the QDKD has been improved to have
double capacity of transmission. In addition, in the original QDKD
and the present improved QDKD, one of the two different secret
keys is discarded to guarantee the complete security of the
encrypted message. This is a serious waste. In quantum
cryptography, generally speaking, the length of the secret key is
shorter than the length of the messages, however, the longer the
secret key, the securer the message. Hence, we suggest to adopt an
additivity strategy to combine the two different keys as one. This
strategy will double the efficiency of generating the secret key
with given length. In fact, such improved QDKD is not secure under
Eve's (the eavesdropper's) intercept-measure-resent attacks
without eavesdropping [5], i.e., Eve intercepts and measures
directly the back photon $t$  and then resends it to Bob, while
Alice and Bob can not detect on the existence of Eve by comparing
their measurement results and it is quite possible that their
secret keys are incorrect. So after the transmission of the secret
key, Alice and Bob have to publicly announce a fraction of their
keys to check whether Eve has ever been in the line during the
transmissions. Incidentally, very similar to the justifications in
Ref.[1,3,4], the security of the present improved QDKD against
eavesdropping can be easily proven. In addition, the original QDKD
[3] is also not secure under Eve's intercept-measure-resent
attacks without eavesdropping [5]. This is a leak of the original
QDKD. We propose to fix it by the public announcement of a
fraction of the secret keys to verify whether Eve has ever been in
the line during the transmission.
 
This work is supported by the NNSF of China under Grant No. 10304022. \\

\noindent[1] K. Bostrom and T. Felbinger, Phys. Rev. Lett.
{\bf89}, 187902 (2002).

\noindent[2] C. H. Bennett and S. J. Wiesner, Phys. Rev. Lett.
{\bf69}, 2881(1992).

\noindent[3] I. P. Degiovanni et al, Phys. Rev. A {\bf 69}, 032310
(2004).

\noindent[4] Q. Y. Cai and B. W. Li, {\it Imporving the capacity
of the Bostrom-Felbinger protocol}, accepted for publication in
Phys. Rev. A .

\noindent[5] Q. Y. Cai, Phys. Rev. Lett. {\bf 91}, 109801 (2003).
\\

\begin{minipage}{360pt}
\begin{center}
\vskip 1cm Table 1.  Corresponding relations among Alice's, Bob's
unitary $u$ operations (i.e., the encoding bits) and Alice's Bell
measurement results on photons $h$ and $t$. Alice's (Bob's)
$u$ operations are listed in the first column (line). \\
\begin{tabular}{ccccc}  \hline
              & $u _0 (00)$           & $u _1 (01)$            & $u _2 (10)$            & $u_3 (11)$
              \\ \hline
$u _0 (00)$   & $|\Psi^+_{ht} \rangle$& $|\Psi^-_{ht} \rangle$ & $|\Phi^+ _{ht}\rangle$ & $|\Phi^-_{ht}\rangle$ \\
$u _1 (01)$   & $|\Psi^-_{ht} \rangle$&$|\Psi^+_{ht} \rangle$ &$|\Phi^-_{ht}\rangle$ &$|\Phi^+_{ht} \rangle$\\
$u _2 (10)$   & $|\Phi^+_{ht} \rangle$& $|\Phi^-_{ht}\rangle$&$|\Psi^+_{ht} \rangle$ & $|\Psi^-_{ht} \rangle$\\
$u_3 (11)$    & $|\Phi^-_{ht}\rangle$&$|\Phi^+_{ht} \rangle$ & $|\Psi^-_{ht} \rangle$& $|\Psi^+_{ht} \rangle$\\
\hline
\end{tabular} \\
\end{center}
\end{minipage}

\end{document}